\documentclass[onecolumn,preprint,showkeys,preprintnumbers,amsmath,amssymb,floatfix]{revtex4}
\usepackage{graphicx}
\usepackage{times}
\begin{document}

\title{Interfacial charge transfer in nanoscale polymer transistors}
\author{Jeffrey H. Worne$^1$, Rajiv Giridharagopal$^1$, Kevin F. Kelly$^1$ and Douglas Natelson$^{1,2}$}

\affiliation{$^1$ Department of Electrical and Computer
Engineering, Rice University, 6100 Main St., Houston, TX 77005}

\affiliation{$^2$ Department of Physics and Astronomy, Rice
University}

\begin{abstract}
Interfacial charge transfer plays an essential role in establishing
the relative alignment of the metal Fermi level and the energy bands
of organic semiconductors.  While the details remain elusive in many
systems, this charge transfer has been inferred in a number of
photoemission experiments.  We present electronic transport
measurements in very short channel ($L < 100$~nm) transistors made
from poly(3-hexylthiophene) (P3HT).  As channel length is reduced, the
evolution of the contact resistance and the zero-gate-voltage
conductance are consistent with such charge transfer.  Short channel
conduction in devices with Pt contacts is greatly enhanced compared to
analogous devices with Au contacts, consistent with charge transfer
expectations.  Alternating current scanning tunneling microscopy
(ACSTM) provides further evidence that holes are transferred from Pt
into P3HT, while much less charge transfer takes place at the Au/P3HT
interface.
\end{abstract}


\maketitle


Understanding the band alignment between organic semiconductors (OSCs)
and metal electrodes is of basic physical interest as well as
significant technological importance\cite{Koch:2008}.  Such energetic
considerations are crucial for optimizing charge injection in organic
light emitting diodes (OLEDs) and organic field-effect transistors
(OFETs).  Similarly, the relative band alignment at such interfaces
also affects the open circuit photovoltage achievable in organic
photovoltaic applications.  Because interfacial dipoles can be used to
modulate the effective work function of a metal surface,
self-assembled monolayers (SAMs) of polar molecules have been used in
both OLEDs and OFETs to engineer charge injection\cite{Campbell:1996,Nuesch:1998,Hamadani:2006}.

Conceptually the issue is straightforward, though very complicated in
detail.  In equilibrium the chemical potential throughout a
metal/organic heterointerface must be constant.  This condition is
achieved through a combination of interfacial dipole formation, charge
transfer, and the self-consistent solution of the electrostatics to
equilibrate carrier drift and diffusion.  Dipole formation and
charge transfer lead to deviations from the Schottky-Mott limit, so
that simple alignment of vacuum levels does not give an accurate
picture of the true heterojunction energetics\cite{Hill:1998}.


The relative alignment of levels is most readily measured via
photoemission experiments.  Recent investigations using
electroluminescent polymers\cite{Tengstedt:2006} as well as small
molecule OSCs and conducting polymers\cite{Koch:2006} have shown Fermi
level pinning at the metal-organic interface.  For the case of holes,
when the Fermi level of the metal is relatively far above the highest
occupied molecular orbital (HOMO) of the molecules (the center of the
disorder-broadened valence band, in the case of polymers), the system
is in the Schottky-Mott limit.  When the Fermi level approaches the
HOMO, strong deviations from the Schottky-Mott limit are
reported\cite{Tengstedt:2006}.  There are several suggested mechanisms
for these deviations, including changes in the interfacial dipole (via
chemical reactions or the ``pillow effect''\cite{Vazquez:2007}),
shifting of frontier orbitals and formation of induced
states\cite{Vazquez:2004}, and charge transfer involving polaron
formation\cite{Tengstedt:2006,Crispin:2006}.  Recent photoemission
experiments examining metal/polymer interfaces show evidence of charge
transfer into valence band tail states and accompanying band
bending\cite{Hwang:2007}.


Other data consistent with charge transfer and band bending at similar
interfaces has been seen in transport
measurements\cite{Hamadani:2005,Hamadani:2006} of contact resistance
in P3HT-based OFETs.  In devices with high electrode work functions,
hole injection remains Ohmic down to very low carrier densities, a
natural result if interfacial charge transfer effectively dopes the
contact interface.  Similarly current-voltage characteristics in
devices with (lower work function) electrode metals giving non-Ohmic
injection are consistent with a nanoscale region (between 10~nm and
100~nm in extent) of reduced mobility at the metal/organic
interface\cite{Burgi:2003,Li:2003,Hamadani:2005b,Ng:2007}.  This may be due to
interfacial structural disorder in the organic semiconductor, but it
would also be compatible with the formation of an effective depletion
region near the contacts due to band bending.  Conduction in P3HT takes
place via hopping through a density of localized states in the valence
band tail.  A local change in the chemical potential farther into
the band tail would simultaneously deplete holes and reduce the 
spatial density of available hopping sites, reducing the mobility as
well.


In this paper we present additional evidence that significant charge
transfer takes place at the Pt/P3HT interface, leading to a population
of mobile holes within a nanoscale distance of the interface.
Measurements of Pt-based OFET device resistance as a function of
channel length, $L$, show a decrease in resistance as $L$ is reduced
below 100~nm.  In the same devices this corresponds to nearly a factor
of 10 increase in zero-gate-bias (ZGB) current over the same $L$
range.  In contrast, identically prepared short channel Au-based
devices show an increase in resistance, as well as ZGB currents orders
of magnitude lower than in Pt devices.  Devices with Au surfaces
functionalized with SAMs known to increase the electrode effective
work function show trends intermediate between the Au and Pt cases.
Further measurements using alternating current scanning tunneling
microscopy (ACSTM) show clear evidence for excess holes in thin P3HT
films on Pt, in marked contrast to ACSTM measurements on identically
prepared P3HT films on Au.

Devices are fabricated on $n+$ silicon with 200nm of thermally grown
oxide as the gate dielectric.  Electrodes are patterned using electron
beam lithography, e-beam evaporation, and liftoff processing.  Using
two-step lithography, various channel lengths, $L$, are fabricated,
from $\sim$50~nm to 5~$\mu$m, with widths, $w$ of 50~$\mu$m for short
channel device studies.  In addition, for comparison a structure of
interdigitated electrodes with $L = 1 - 50$~$\mu$m and $w = 200~\mu$m
is fabricated on the same chip in order to allow standard transmission
line measurements of P3HT field effect mobility.  Electrode films are
15~nm of either gold or platinum with a 1~nm titanium adhesion layer.
Following lithography and liftoff, the devices are cleaned under
ultraviolet exposure, 1 minute of oxygen plasma, and a 1 minute soak
in piranha etch (3:1 H$_{2}$SO$_{4}$:H$_{2}$O$_{2}$ (30\%)).  For
functionalized Au electrode devices, the piranha etch step is followed
by self-assembly of fluorinated oligo(phenylene ethynylene)
(F-OPE)\cite{Hamadani:2006} by immersion for 24 h in a solution of
F-OPE at 0.25~mg/mL concentration in 1:1 ethanol:chloroform under
nitrogen gas, with standard thioacetate deprotection
chemistry\cite{Cai:2002}.

After the SiO$_{2}$ dielectric surface is treated with
octadecyltricholorosilane (OTS) (10 $\mu$L OTS in 8~mL hexadecane,
assembled in the dark at room temperature for $\sim$12 hours), P3HT
(0.1\% by weight in chloroform) is then spin-coated to a thickness of
5~nm (measured via AFM) over the entire substrate.  Contact to the
substrate is made, and the Si becomes the gate of the device.  Figure 1b
illustrates a typical short channel device.  Devices are transferred
to a variable temperature vacuum probe station immediately upon
completion of spin coating.  Devices remain in vacuum ($\sim
10^{-6}$~mTorr) for at least 1 hour before any measurements are
performed.  Individual OFETs are isolated from one another by using a
probe tip to scratch away the P3HT film in a rectangle around each
device to prevent stray current paths.

Using the interdigitated electrodes of fixed $w$ but varying $L$,
device characteristics ($I_{\mathrm D}$ vs. $V_{\mathrm{SD}}$ at fixed
$V_{\mathrm{G}}$, where $I_{\mathrm{D}}$ is drain current and
$V_{\mathrm{SD}}$ is source-drain voltage) are measured via a
semiconductor parameter analyzer.  In the shallow channel regime
($|V_{\mathrm{SD}}|<<|V_{\mathrm{G}}|$) we find these characteristics
to be linear, consistent with Ohmic injection.  As in our previous
work\cite{Hamadani:2004,Hamadani:2006} we use the transmission line
approach to characterize the contact resistances and field-effect
mobilities as a function of gate voltage, $V_{\mathrm{G}}$.  In the
shallow channel regime at a particular $V_{\mathrm{G}}$, the measured
source-drain resistance $R = (\partial I_{\mathrm{D}}/\partial
V_{\mathrm{SD}})^{-1}$ is plotted as a function of $L$ and found to be
linear for long ($L > 1~\mu$m) devices.  Channel resistance,
$R_{\mathrm{C}}$, is determined by the slope of this line and the
effective contact resistance, $R_{\mathrm{S}}$ is determined by
extrapolating this line to zero channel length, as shown in Fig.~1a.
This linearity observed in longer devices indicates that the overall
film uniformity is good, even for the relatively thin P3HT layers used
here.

This extrapolation to infer $R_{\mathrm{S}}$ assumes that the device
properties are uniform down to arbitrarily short channel lengths.  In
practice this may not be true, for a number of reasons.  For example,
a change in polymer morphology near the metal contacts would lead to
deviations from linearity in $R(L)$ at short distances as contact
regions interact with each other\cite{Gundlach:2008}.  Similarly,
there is indirect evidence\cite{Burgi:2003,Li:2003,Hamadani:2005b} that
nanoscale regions of poor effective mobility can exist near contacts
due to band bending and depletion effects.  Note that the characteristic
size of these regions is constrained experimentally\cite{Hamadani:2005b}
to less than $\sim 100$~nm, since larger regions would be detectable
within the resolution of scanning potentiometry experiments.  If
two such depletion regions were to intersect for sufficiently short
channel devices, one would expect the measured $R(L)$ to have
an up-turn as $L \rightarrow 0$, as shown by the red line in 
Fig. 1a.

Figure 2 shows these $R$ vs. $L$ plots for three sets of Au, F-OPE/Au,
and Pt devices at 300~K and $V_{\mathrm{G}} = -70$~V for
\textit{shorter} channel devices.  The resistance was measured via
$V_{\mathrm{SD}}$ sweeps from 0 to -500~mV, while gate voltages from
0~V to -70~V (in 10~V steps) were examined.  The mobilities at
$V_{\mathrm G}~=~-70$~V, inferred from the length dependences, were
$5.3 \times 10^{-2}$~cm$^{2}$/Vs (Au), $5.3 \times
10^{-2}$~cm$^{2}$/Vs (F-OPE/Au), and $8.7\times 10^{-2}$~cm$^{2}$/Vs
(Pt).  Initial measured resistances from the 5~$\mu$m channel length
devices span roughly an order of magnitude between the three different
electrode materials, with the highest resistance coming from the gold
electrodes and the lowest resistance from the platinum devices.  As
the channel length is reduced significantly below 1~$\mu$m, there are
qualitative differences between the three types of devices.  The
Au-based devices have much-increased resistances, while the Pt-based
devices have decreased resistances.  The F-OPE/Au structures are
intermediate in their small $L$ properties.  We analyzed several
ensembles of identically prepared devices, and all exhibited these
trends.  These trends continued to hold at other gate voltages and as
$T$ was reduced.  Channel lengths in the short-channel devices were
measured via electron microscopy \textit{after} electrical
characterization, to ensure that electrode contamination during
imaging did not affect the results.

At the smallest channel lengths there is significant device-to-device
variability in $R$, much more so than in the longer channel devices.
This appears to reflect that microscopic differences in the
metal/organic interface can have significant local impact on the
injection process and contact resistances, even when large-scale
properties are uniform and well defined.  However, even with this
variability the overall trends and the systematic deviation between
the different electrode types are clear.

One possibility is that the morphology of the P3HT may differ at the
metal/P3HT interface for the three different electrode materials,
leading to varying contact properties.  To test this, we performed
tapping-mode atomic force microscopy (AFM) scans of a 3~$\mu$m
$\times$ 3~$\mu$m area that encompassed the electrode/P3HT interface
for both platinum and gold electrodes.  The results are shown in
Figure 3.  The topographic images show that the polymer film on the
oxide adjacent to the electrode edge is smooth (rms roughness of 0.83
nm for the P3HT on oxide next to Au; 0.78~nm for the P3HT on oxide
next to Pt) and largely featureless for both electrode materials.
Using phase imaging, more detail is observable, with some indications
of the fibril morphology sometimes observed\cite{Merlo:2003} in P3HT.
However, there is little difference between the two images, at least
down to the resolution of our microscope.  This implies that gross
morphology changes near the contacts are not responsible for the
difference in contact resistance properties.

We suggest instead that the explanation for the difference lies in
interfacial charge transfer between the electrodes and the P3HT due to
the energetics of band alignment.  Previous
experiments\cite{Rep:2003,Hamadani:2005,Hamadani:2006} have suggested
that in the absence of unintentional doping from the environment the
alignment between the Au Fermi level and the P3HT HOMO is such that a
significant hole injection barrier exists.  In this case one may
expect the tail of the P3HT valence band to be locally depleted in the
vicinity of the of the interface.  In contrast, measurements involving
the interface between P3HT and higher work function systems such as
Pt\cite{Hamadani:2005,Hamadani:2006} and F-OPE/Au\cite{Hamadani:2006}
show much lower contact resistances and persistent Ohmic injection in
field-effect structures even under treatment conditions where the bulk
two-terminal conductivity of the P3HT is immeasurably small.
Consistent with recent photoemission experiments\cite{Hwang:2007}, 
we suggest that the charge transfer responsible for pinning the
Pt Fermi level above the P3HT HOMO populates the tail states with
mobile holes, leading to comparatively enhanced conduction in
the shortest devices.

This interpretation is supported by other experimental signatures.
First, we examine the low $V_{\mathrm{SD}}$ transistor characteristics
as a function of $V_{\mathrm{G}}$ for short-channel devices with {\em
  identical geometries}.  This is more revealing than
$I_{\mathrm{D}}-V_{\mathrm{G}}$ transfer characteristics since the
changing channel geometry as $L$ is reduced, high electric fields at
large $V_{\mathrm{SD}}$, and hysteresis in $V_{\mathrm{G}}$ complicate
the interpretation of inferred threshold voltages.  A typical result
is shown in Fig. 4, for Pt (top) and Au (bottom) devices with $L
\approx$~100~nm at $T = 300$~K.  Note that there is a large
qualitative difference between these devices.  The Pt device shows
significant background conduction at $V_{\mathrm{G}}=0$, qualitatively
similar to what is seen in the presence of doping.  This conduction
vanishes in adjacent, simultaneously fabricated Pt devices as channel
lengths are increased into the micron range, showing that it is
an effect of the contacts.

We investigate this further by examining the two-terminal source-drain
conductance at $V_{\mathrm{G}}=0$ for the short channel devices ($L <
5~\mu$m) at $V_{\mathrm{SD}}= -0.5$~V.  Representative data are shown
in Fig. 5.  At the 5~$\mu$m limit the ZGB current is essentially the
same for all devices.  As the channel length decreases, however, the
measured currents in devices with platinum contacts continue to
increase by orders of magnitude, while the measured currents in
devices with gold contacts do not follow this trend.  Rather, the Au
device $V_{\mathrm{G}}=0$ currents tend to be fairly constant over the
measured channel length range.  The current measured from the F-OPE/Au
electrodes remains fairly constant and similar to that of gold over
the longer channel lengths, but as the channel lengths decrease well
below 200~nm, the zero-bias current displays a rapid increase.  Since
all of the devices of a given electrode composition are prepared
simultaneously and measured in one run, these dependences of the
$V_{\mathrm{G}}=0$ conduction on $L$ {\em cannot} arise from dopants from the
environment.  

To further examine charge transfer between the electrodes and the
P3HT film, we used an extension of scanning tunneling microscopy (STM)
known as alternating current scanning tunneling microscopy
(ACSTM).  STM and related spectroscopic techniques have long been used
to characterize the electronic properties of conducting polymer
materials.  More common spectroscopies provide high-resolution analysis of
properties such as the density of states and work function, yet fail to provide
capacitance versus voltage ($C-V$) information, which is critical for
understanding dopant effects. Unlike conventional STM, microwave
frequency ACSTM can be used to obtain local $C-V$ data.

In the ACSTM technique, microwave frequency radiation is applied to
the tip-sample junction.  By exploiting the nonlinear behavior of the
junction using high frequencies, ACSTM can be used to acquire
capacitance information.  Specifically, the tunneling current response
at microwave frequencies when imaging semiconductors reflects the
$\partial C/\partial V$ data\cite{Bourgoin:1994}. In this way, ACSTM
can provide high-resolution spatial capacitance information on
semiconducting polymer films.  The magnitude of the ACSTM signal
ultimately reflects the carrier concentration in the substrate.  For
example, this microwave signal has been used previously in both STM
and AFM mode to measure the dopant concentration in
silicon\cite{McCarty:2001,Kelly:2005,Schmidt:1999}.

For this ACSTM experiment a loop antenna geometry was used to apply
microwave radiation to the junction similar to studies previously
reported\cite{Lee:2005}.  We used the difference frequency ACSTM
technique\cite{McCarty:2001,Kelly:2005,Schmidt:1999,Bumm:1996} to
simplify detection of the high frequency modulation of the tunneling
signal by allowing the use of conventional lock-in amplifier
equipment.  Care was taken to ensure that the ACSTM data on both
samples were acquired under nearly identical scanning
conditions\cite{note}.  The ACSTM signal was acquired at each point
during scanning, thus producing $\partial C/\partial V$ spectral
images (magnitude and phase) and the topography image simultaneously.
All scans were taken in ultrahigh vacuum at base pressures on the
order of 10$^{-10}$ torr using a commercial RHK STM and mechanically
cut Pt:Rh (80:20) tips.  The STM was modified to include the antenna
so that the loop antenna encircled the STM tip during scanning,
similar to a method reported previously\cite{Lee:2005} and could be
moved aside for tip and sample exchange.  P3HT films were spin-coated
onto Au- and Pt-coated SiO$_{2}$/Si substrates following the same
procedure as in the fabrication of the OFET devices.

In Fig. 6, the topography and ACSTM image are both shown at various
bias conditions for P3HT deposited on Au and Pt,
respectively\cite{note2}.  The ACSTM images are shown at the same
relative color scale in units of mV, directly proportional to local
$\partial C/\partial V$ and therefore carrier concentration (as seen
in doped Si devices\cite{McCarty:2001}).  Bear in mind that the
carrier concentration seen in semiconductors in ACSTM depends to some
degree on the bias conditions and gap geometry.  The gap resistance
was kept constant at $\sim 100~\mathrm{G}\Omega$ to eliminate the
influence of tip height on the resulting data.  A direct comparison of
the ACSTM data on Au and Pt shows a marked contrast in carrier
concentration.  Averaged over the scan area, the carrier concentration
is 25\% larger for P3HT on Pt than on Au, but there is also a higher
spatial variance on the Pt sample despite the comparable surface
structure.  This is consistent with the interpretation that there are
more mobile holes in the P3HT layer on Pt.  Measurements taken at
multiple locations on each film confirm these results.  In addition,
the percentage difference between average ACSTM signal on Pt vs. Au
increases to 69\% as the tip bias relative to the sample is increased
to 2~V.  This shows that the carriers present on Au are easier to
deplete relative to those on Pt.  It is difficult to compare
quantitatively the ACSTM data in these vertical layered structures
with the situation in the transistor geometry.  However, the ACSTM
data presented here clearly indicate that Pt increases the carrier
concentration in the P3HT film more than the Au layer.

Interestingly, both films show inhomogeneities in the magnitude and
phase of their AC response that do not correlate in any immediately
obvious way with sample topography, although Figs. 6A and 6C show that
the surface features of the two films are similar.  In particular the
variation takes place on length scales that are not correlated with
intrinsic structural properties of the underlying polycrystalline
metal films, such as grain size.

The transport and ACSTM indications of charge transfer are consistent
with simple calculations using a model for charge transfer developed
by Paasch and Scheinert\cite{Paasch:2007}.  Their model was developed
to explain charge transfer and band bending in thin layers of
disordered organic semiconductor on top of metal films.  This is
precisely the situation of the ACSTM samples, and at zero source-drain
bias should be reasonably applicable to the organic channel in the
immediate vicinity of the source or drain electrodes.  Assuming an
exponential density of states for the organic semiconductor
characterized by an energy scale $k_{\mathrm B}T_{0} = 0.1$~eV, the
local potential in the organic semiconductor of layer thickness $d$ as
a function of distance $x$ away from the metal interface is given by
\begin{equation}
U(x)= U_d \pm {k_{\mathrm{B}}T_{0}}\ln \left[1 + \tan^{2}\left(\frac{d-x}{2L_{d}}\exp \frac{|U_d|}{2 k_{\mathrm{B}}T_{0}} \right)\right],
\label{eq:paasch}
\end{equation}
where $U_d$ is the potential at $x=d$, $L_d \equiv \sqrt{\epsilon
  \epsilon_0 k_{\mathrm{B}}T/2e^2n}$ is the intrinsic screening
length, and $n = \sqrt{N_cN_v}\exp(-E_g/2k_{\mathrm{B}}T_{0})$ is the
effective intrinsic carrier density\cite{Paasch:2007}.  Here
$\epsilon$ is the relative dielectric constant of the organic
semiconductor, $E_g$ is the band gap of the organic semiconductor,
$N_c$ and $N_v$ are the effective densities of states in the
conduction and valence bands.  The $\pm$ sign is positive (negative)
for hole(electron) accumulation.  The Poisson equation can be used to
find the relationship between $d$ and $U_d$:
\begin{equation}
d = 2 L_{d} \exp \left(- \frac{|U_{d}|}{2 k_{\mathrm{B}}T_{0}}\right)\arctan \left[\exp \left(\frac{|U_s-U_d|}{k_{\mathrm{B}}T_{0}}\right)-1 \right]^{0.5}.
\label{eq:paasch2}
\end{equation}
Here $U_s \equiv E_{g}/2 - \Phi_{h}$, where $\Phi_{h}$ is the hole
injection barrier.  Rough values used for our estimations are $N_c,~N_v =
10^{21}$~cm$^{-3}$, $E_g = 2$~eV, $\epsilon = 3.24$, and $T = 300$~K.

Working with Eqs.~(\ref{eq:paasch}, \ref{eq:paasch2}) and Poisson's
equation, we can determine the charge density in the organic
semiconductor layer.  Varying the metals involved affects the local
potential $U(x)$ and carrier density by altering $\Phi_h$.  Recalling
previous photoemission results for P3HT on platinum and
gold\cite{Hamadani:2005}, for platinum with $\Phi_{h} = 0.6$~eV, the
charge carrier density very near the interface is estimated to be
$\sim 9.6\times10^{17}$cm$^{-3}$; for gold with $\Phi_{h} = 1.2$~eV,
we find $\sim -10^{18}$~cm$^{-3}$, with the negative sign indicating a
\textit{depletion} of hole density.  These precise numbers should be
viewed cautiously, since different surface preparations can alter the
effective work functions of metals and the density of states values
used are rough.  The trend, however, from Au to Pt should be robust
and these results are unlikely to change qualitatively upon inclusion
of corrections for a Gaussian density of states for the polymer.  For
both the Pt, the $\Phi_{h} = 0.6$~eV value is consistent with prior
photoemission experiments\cite{Hamadani:2005} as well as recent Fermi
level pinning results for P3HT and estimates of polaron formation
energy relevant to Pt/P3HT charge transfer\cite{Tengstedt:2006}.  The
estimate of $\Phi_{h}$ for Au also comes from the same photoemission
experiments\cite{Hamadani:2005}.  The resulting calculation supports
the claim that higher work function metals, with Fermi levels pinned
relatively close to the valence band, locally dope the organic
material very near the interface.

We have presented a series of experiments examining the contact
resistance and ZGB current in very short channel bottom-contact OFETs
incorporating P3HT.  The striking differences observed between devices
with Pt and Au electrodes are consistent with expectations of
metal/P3HT charge transfer inferred from photoemission experiments.
In particular, it appears that a significant density of mobile holes
are transferred from Pt into neighboring P3HT on the nanoscale.  ACSTM
measurements further support this conclusion, showing a significantly
higher free carrier response in P3HT films on Pt compared to
identically prepared films on Au.  Further experiments, particularly
those probing spatial scales comparable to small numbers of polymer
chains, should be able to shed further light on the complex problem of
charge transfer and Fermi level pinning, complementing spatially
averaging techniques such as photoemission.  Techniques like ACSTM
point the way toward detailed quantitative assessment of local charge
density, information that will be extremely useful in refining
theoretical models of these important but complicated problems.

\begin{acknowledgments}
The authors gratefully acknowledge Jun Zhang for experimental
assistance, Paul Weiss for useful discussions, Prof. J. W. Ciszek and
Prof. J. M. Tour for synthesis of the F-OPE molecule, and the support
from NSF grant ECCS-0601303.  RG acknowledges the support of an NSF
graduate fellowship.  DN also aknowledges the David and Lucille
Packard Foundation, the Alfred P. Sloan Foundation, the Robert
A. Welch Foundation, and the Research Corporation.  KFK also
acknowledges the Rochester MURI on Nanoscale Subsurface Spectroscopy
and Tomography (F49620-031-0379), administered by the Air Force Office
of Scientific Research.
\end{acknowledgments}

\clearpage

\begin{figure}[!h]
\begin{center}
\includegraphics[clip, width=8.5cm]{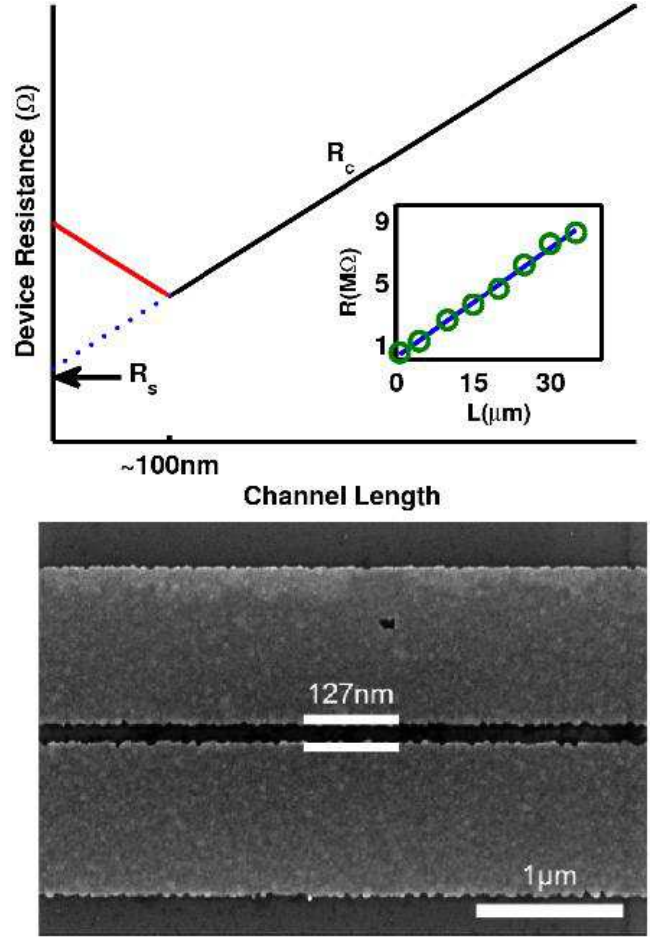}
\end{center}
\caption{Top:  Expected two-terminal resistance as a function of channel length,
with deviations at very short channels expected if contact interface
regions have properties that differ significantly from the bulk.  Inset: Data on a series of devices with channel width 200~$\mu$m at $V_{\mathrm G}=-50$~V and $T = 300$~K, $V_{\mathrm{SD}}$ from 0 to -1~V, showing the long-channel dependence.  Of interest in our experiments is the very short channel limit.
Bottom:  Electron micrograph of the center of a typical short channel device fabricated by electron beam lithography.} 
\label{fig1}
\end{figure}

\clearpage

\begin{figure}[!h]
\begin{center}
\includegraphics[clip, width=8.5cm]{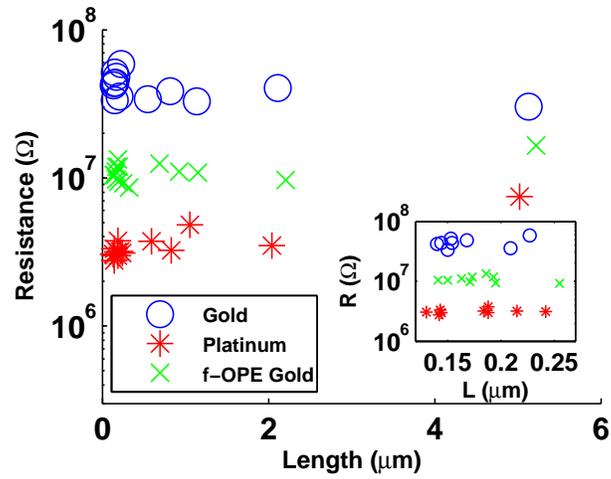}
\end{center}
\caption{Two-terminal resistance at low $V_{\mathrm{SD}}$ at 300~K and $V_{G}=-70$~V for various series of OFETs incorporating different electrode materials.  Note the greatly enhanced resistance in very short channel Au devices compared with Pt devices.  Inset shows a more clear view of the short channel data.} \label{fig2}
\end{figure}

\clearpage

\begin{figure}[!h]
\begin{center}
\includegraphics[clip, width=8.5cm]{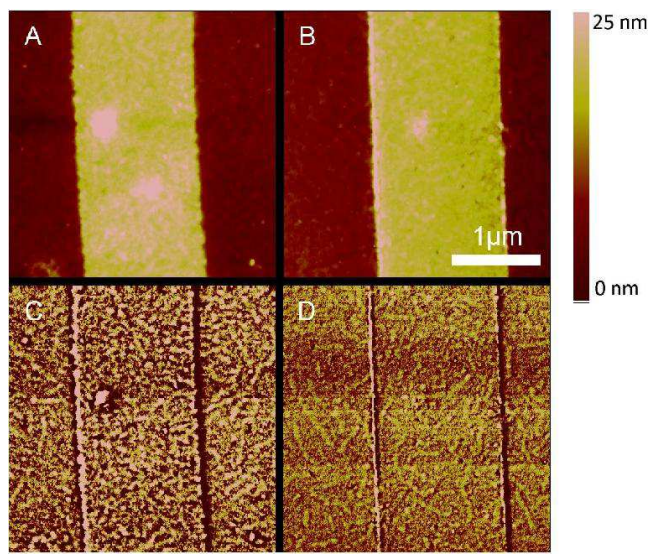}
\end{center}
\caption{Atomic force microscopy images of P3HT films in contact with gold (A, C) and platinum (B, D) electrodes.  There is no significant difference in P3HT morphology at the two interfaces, suggesting that the observed differences in electrical properties does not result from greatly varying amounts of disorder.} \label{fig3}
\end{figure}

\clearpage

\begin{figure}[!h]
\begin{center}
\includegraphics[clip, width=8.5cm]{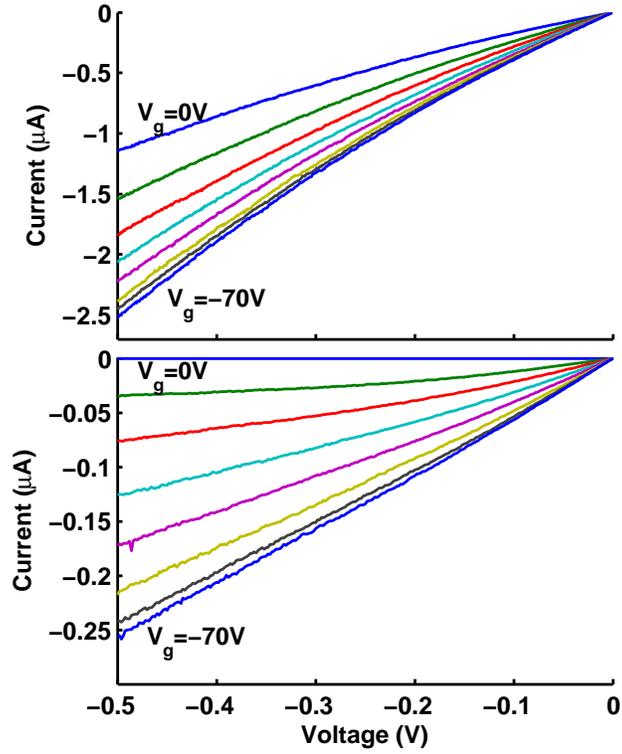}
\end{center}
\caption{Transistor characteristics at low $V_{\mathrm{SD}}$ for identically prepared Pt (top) and Au (bottom) devices with channel lengths $\approx$144~nm , width 50~$\mu$m at $T = 300$~K.  These devices have essentially identical geometries, yet the Pt-based device shows significant background conduction even at $V_{\mathrm{G}} = 0$.  This qualitative difference vanishes as channel lengths exceed the few micron scale.} \label{fig4}
\end{figure}

\clearpage

\begin{figure}[!h]
\begin{center}
\includegraphics[clip, width=8.5cm]{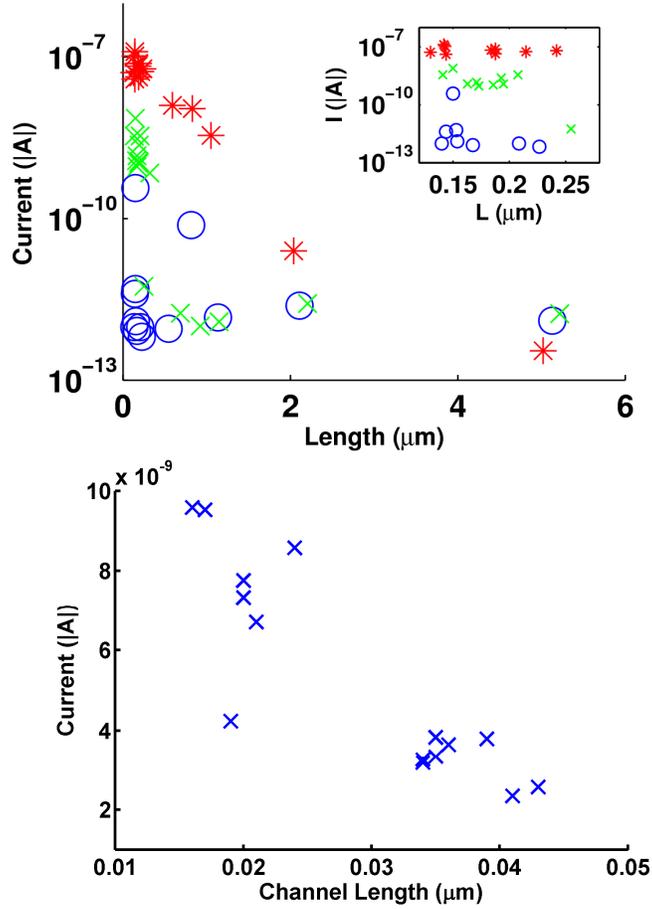}
\end{center}
\caption{Top:  ZGB current (measured at $V_{\mathrm{G}}=0$, $V_{\mathrm{SD}}=-0.5$~V) at 300~K for the same series of devices (same symbols) in Fig. 2.  Notice the great enhancement in ZGB current for extremely short channel Pt and FOPE-Au devices, much larger than simple geometric expectations would suggest.  Bottom:  Data acquired in a separately fabricated ensemble of Pt-based devices made using the nanogap method of Ref.~\protect{\cite{Fursina:2008}}.} \label{fig5}
\end{figure}

\clearpage

\begin{figure}[!h]
\begin{center}
\includegraphics[clip, width=8.5cm]{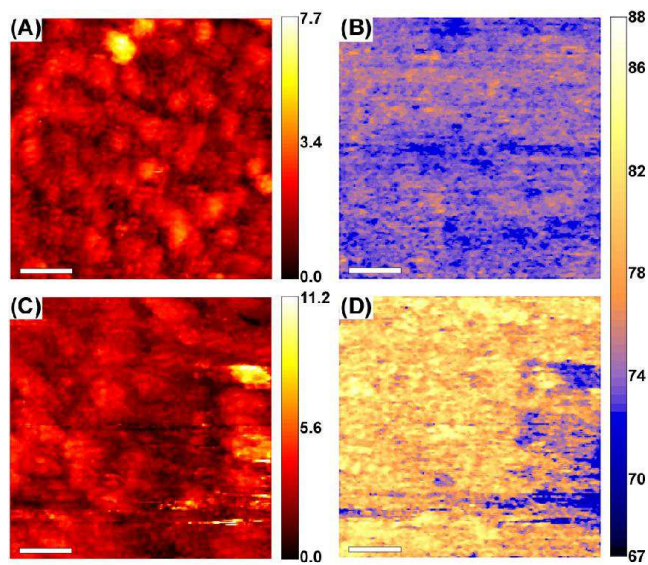}
\end{center}
\caption{Topography (A, C) and ACSTM (B, D) magnitude images of P3HT films deposited on Au (A, B) and Pt (C, D), respectively.   The simultaneously acquired ACSTM images are shown at the same relative color scale.  All images were acquired at -1.0 V sample bias, 10 pA tunneling current, and are 473 $\times$ 473~nm$^{2}$.  The scalebars on all images are 94.6 nm. The color bars are in units of nm for the topography images and mV for the ACSTM images.  The greater magnitude in (D) indicates an excess of mobile holes for the P3HT/Pt case relative to the P3HT/Au case of (B).} \label{fig6}
\end{figure}





\clearpage


\begin{thebibliography}{99}


\bibitem{Koch:2008}
Koch, N.  Energy levels at interfaces between metals and conjugated organic molecules. {\it J. Phys: Condens. Matter} 2008, 20, 184008.

\bibitem{Campbell:1996}
Campbell, I. H.; Rubin, S.; Zawodzinski, T. A.; Kress, J. D.; Martin, R. L.; Smith, D. L.; Barashkov, N. N.; Ferraris, J. P. Controlling Schottky energy barriers in organic electronic devices using self-assembled monolayers. {\it Phys. Rev. B} 1996, 54, R14321-R14324.

\bibitem{Nuesch:1998}
N{\"u}esch, F.; Rotzinger, F.; Si-Ahmed, L.; Zuppiroli, L. Chemical potential shifts at organic device electrodes produced by grafted monolayers.  {\it Chem. Phys. Lett.} 1998, 288, 861-867.

\bibitem{Hamadani:2006}
Hamadani, B. H.; Corley, D. A.; Ciszek, J. W.; Tour, J. M.; Natelson, D. Controlling charge injection in organic field-effect transistors using self-assembled monolayers.  {\it Nano Lett.} 2006, 6, 1303-1306.


\bibitem{Hill:1998}
Hill, I. G.; Rajagopal, A.; Kahn, A.; Hu, Y. Molecular level alignment at organic semiconductor-metal interfaces.  {\it Appl. Phys. Lett.} 1998, 73, 662-664. 

\bibitem{Tengstedt:2006}
Tengstedt, C.; Osikowicz, W.; Salaneck, W. R.; Parker, I. D.; Hsu C.-H.; Fahlman, M. Fermi-level pinning at conjugated polymer interfaces.  {\it Appl. Phys. Lett.} 2006, 88, 053502.

\bibitem{Koch:2006}
Koch N.;  Vollmer, A.  Electrode-molecular semiconductor contacts:  Work-function-dependent injection barriers versus Fermi-level pinning.  {\it Appl. Phys. Lett.} 2006, 89, 162107.

\bibitem{Vazquez:2007}
V{\'a}zquez, H.; Dappe, Y. J.; Ortega, J.; Flores, F. Energy level alignment at metal/organic semiconductor interfaces:  ``Pillow'' effect, induced density of interface states, and charge neutrality level.  {\it J. Chem. Phys.} 2007, 126, 144703.

\bibitem{Vazquez:2004}
V{\'a}zquez, H.; Flores, F.; Oszwaldowski, R.; Ortega, J.; Perez, R.; Kahn, A.  Barrier formation at metal-organic interfaces:  dipole formation and the charge neutrality level.  {\it Appl. Surf. Sci.} 2004, 234, 1-4.

\bibitem{Crispin:2006}
Crispin, A.; Crispin, X.; Fahlman, M.; Berggren, M.; Salaneck, W. R. Transition between energy level alignment regimes at a low band gap polymer-electrode interfaces.  {\it Appl. Phys. Lett.} 2006, 89, 213503.

\bibitem{Hwang:2007}
Hwang, J.; Kim, E.-G.; Liu, J.; Br{\'e}das, J.-L.; Duggal, A.; Kahn, A. Photoelectron spectroscopic study of the electronic band structure of polyfluorene and fluorene-arylamine copolymers at interfaces.  {\it J. Phys. Chem. C} 2007, 111, 1378-1384.

\bibitem{Burgi:2003}
B{\"u}rgi, L.; Richards, T. J.; Friend, R. H.; Sirringhaus, H. Close look at charge carrier injection in polymer field-effect transistors.  {\it J. Appl. Phys.} 2003, 94, 6129-6137.

\bibitem{Li:2003}
Li, T.; Ruden, P. P.; Campbell, I. H.; Smith, D. L. Investigation of bottom-contact organic field effect transistors by two-dimensional device modeling. {\it J. Appl. Phys.} 2003,  93, 4017-4022.

\bibitem{Hamadani:2005b}
Hamadani, B. H.; Natelson, D. Nonlinear charge injection in organic field-effect transistors.  {\it J. Appl. Phys.} 2005, 97, 064508.

\bibitem{Ng:2007}
Ng, T. N.; Silveira, W. R.; Marohn, J. A.  Dependence of charge injection on temperature, electric field, and energetic disorder in an organic semiconductor.  {\it Phys. Rev. Lett.} 2007, 98, 066101. 

\bibitem{Hamadani:2005}
Hamadani, B. H.; Ding, H.; Gao, Y.; Natelson, D.  Doping-dependent charge injection and band alignment in organic field-effect transistors.  {\it Phys. Rev. B} 2005, 72, 235302.

\bibitem{Cai:2002}
Cai, L. ; Yao, Y.; Yang, J.; Price, Jr., D. W.; Tour, J. M. Chemical and potential assisted assembly of thioacetyl-terminated oligo(phenylene ethynylene)s on gold surfaces.  {\it Chem. Mater.} 2002, 14, 2905-2909.

\bibitem{Hamadani:2004}
Hamadani, B. H.; Natelson, D.  Temperature-dependent contact resistances in high quality polymer field-effect transistors.  {\it Appl. Phys. Lett.} 2004, 84, 443-445.

\bibitem{Gundlach:2008}
Gundlach, D. J.; Royer, J. E.; Park, S. K.; Subramanian, S.; Jurchescu, O. D.;
Hamadani, B. H.; Moad, A. J.; Kline, R. J.; League, L. C.; Kirillov, O.; 
Richter, C. A.; Kushmerick, J. G.; Richter, L. J.; Parkin, S. R.; Jackson, T. N.; Anthony, J. E.  Contact-induced crystallinity for high-performance soluble acene-based transistors and circuits. {\it Nature Mater.} 2008, 7, 216-221.


\bibitem{Merlo:2003}
Merlo, J. A.; Frisbie, C. D. Field effect conductance of conducting polymer nanofibers.  {\it J. Polymer Sci. B:  Polymer Phys.} 2003, 41, 2674-2680. 

\bibitem{Rep:2003}
Rep, D. B. A.; Morpurgo, A. F.; Klapwijk, T. Doping-dependent charge injection into regioregular poly(3-hexylthiophene).  {\it Org. Elect.} 2003, 4, 201-207.

\bibitem{Bourgoin:1994}
Bourgoin, J. P.; Johnson, M. B.; Michel, B. Semiconductor characterization with the scanning surface harmonic microscope.  {\it Appl. Phys. Lett.} 1994, 65, 2045-2047.

\bibitem{McCarty:2001}
Donhauser, Z. J.; McCarty, G. S.; Bumm, L. A.; Weiss P. S. High resolution dopant profiling using a tunable AC scanning tunneling microscope.  In {\it Characterization and Metrology for ULSI Technology: 2000 International Conference}, Seiler, D. G. {\it et al.}, Eds.; American Institute of Physics: New York, 2001, pp. 641-646.

\bibitem{Kelly:2005}
Kelly, K. F.; Donhauser, Z. J.; Mantooth, B. A.; Weiss, P. S. Expanding the capabilities of the scanning tunneling microscope.  In {\it NATO ASI Series II: Mathematics, Physics, and Chemistry}, Vilarinho, P.; Rosenwaks, Y.; Kingon, A., Eds.; Springer: New York, 2005, pp 153-171.

\bibitem{Schmidt:1999}
Schmidt, J.; Rapoport, D. H.; Fr{\"o}hlich, H.-J. Microwave-frequency alternating current scanning tunneling microscopy by difference frequency detection:  Atomic resolution imaging on graphite.  {\it Rev. Sci. Instr.} 1999, 70, 3377-3380.

\bibitem{Bumm:1996}
Bumm, L. A.;  Arnold, J. J.;  Cygan, M. T.; Dunbar, T. D.; Burgin, T. P. ; Jones, II, L.; Allara, D. L.; Tour, J. M.; Weiss, P. S. Are single molecular wires conducting?  {\it Science} 1996,  271, 1705-1707.

\bibitem{Lee:2005}
Lee, J.; Tu, X.; Ho, W. Spectroscopy and microscopy of spin-sensitive rectification current inducted by microwave radiation.  {\it Nano Lett.} 2005, 5, 2613-2617.

\bibitem{note} Frequencies used were 300 and 300.004~MHz,
  approximately 3.7~dbm for the Pt sample and 3.0~dbm for the Au
  sample. The difference is due to the change in antenna position when
  switching between samples, and was adjusted by monitoring the input
  at the lock-in amplifier to ensure that the initial out-of-tunneling
  input signal strength at the difference frequency was the same for
  both samples. The same tip was used for both samples.

\bibitem{note2}
Topography images were processed using plane-/offset-subtraction and median filtering in MATLAB, while ACSTM images were only median filtered.

\bibitem{Paasch:2007}
Paasch, G.; Scheinert, S. Space charge layers in organic field-effect transistors with Gaussian or exponential semiconductor density of states.  {\it J. Appl. Phys.} 2007, 101, 024514.


\bibitem{Fursina:2008}
Fursina, A.; Lee, S.; Sofin, R. G. S.; Shvets, I. V.; Natelson, D.  Nanogaps with very large aspect ratios for electrical measurements.  {\it Appl. Phys. Lett.} 2008, 92, 113102.











\end{thebibliography}
\end{document}